\documentclass[
preprint,
3p,twocolumn
]{elsarticle}

\usepackage{hyperref}

\usepackage{graphicx}
\usepackage{dcolumn}
\usepackage{bm}
\usepackage[caption=false]{subfig}
\usepackage{amsmath}
\usepackage[hyperref]{xcolor}

\hypersetup{colorlinks=true}

\journal{Computational Material Science}

\biboptions{sort&compress}

\begin{document}

\begin{frontmatter}

\title{A general computational method for electron emission and thermal effects in field emitting nanotips}

\author[mymainaddress]{A. Kyritsakis\corref{mycorrespondingauthor}}
\ead{andreas.kyritsakis@helsinki.fi, akyritsos1@gmail.com}

\author[mymainaddress]{F. Djurabekova}
\cortext[mycorrespondingauthor]{Corresponding author}

\address[mymainaddress]{Deparment of Physics and Helsinki Institute of Physics, University of Helsinki, PO Box 43 (Pietari Kalmin katu 2), 00014 Helsinki, Finland}

\begin{abstract}
Electron emission from nanometric size emitters becomes of increasing interest due to its involvement to sharp electron sources, vacuum breakdown phenomena and various other vacuum nanoelectronics applications. The most commonly used theoretical tools for the calculation of electron emission are still nowadays the Fowler-Nordheim and the Richardson-Laue-Dushman equations although it has been shown since the 1990's that they are inadequate for nanometrically sharp emitters or in the intermediate thermal-field regime. In this paper we develop a computational method for the calculation of emission currents and Nottingham heat, which automatically distinguishes among different emission regimes, and implements the appropriate calculation method for each. Our method covers all electron emission regimes (thermal, field and intermediate), aiming to maximize the calculation accuracy while minimizing the computational time. As an example, we implemented it in atomistic simulations of the thermal evolution of Cu nanotips under strong electric fields and found that the predicted behaviour of such nanotips by the developed technique differs significantly from estimations obtained based on the Fowler-Nordheim equation. Finally, we show that our tool can be also successfully applied in the analysis of experimental $I-V$ data.  
\end{abstract}

\begin{keyword}
	\texttt{electron emission, tunnelling, sharp emitters, nanotips}
\end{keyword}

\end{frontmatter}


\section{Introduction}

The accurate quantitative estimation of field electron emission may play a crucial role in the design of nanoelectronic devices and other technological applications with requirements of very high precision. Moreover, the behaviour of metal surfaces prior to vacuum arcing observed at very high electric fields is closely associated with the dynamics of emission currents growing intensively just before the event takes place. 

Traditionally, the different regimes of electron emission: thermionic, field and photoemission, are considered separately, since they all appear at different conditions and can be exploited independently. They also clearly follow different laws introduced to describe each regime since 1920s: the Richardson-Laue-Dushman (RLD) law for thermionic emission \cite{Richardson, Dushman} and the Fowler-Nordheim (FN) equation for field emission \cite{FN1928, Nordheim1928}. Nevertheless, Jensen already showed that thermionic and field emission cannot be always separated and developed the General Thermal-Field (GTF) theory \cite{Jensen2006, Jensen2007, Jensen2008} providing with reasonable approximations for all regimes (thermionic, field and intermediate).  This is particularly true in the violent condition of vacuum arcing where both field and thermionic emission cannot be neglected \cite{Anders}. 

Meanwhile, it has been shown since the 1990's \cite{CutlerAPL} that the usage of the classical Schottky-Nordheim (SN) \cite{Nordheim1928, Schottky1923} model for the tunnelling barrier -- used by both FN and GTF theories -- is inadequate for sharp emitters with radii of curvature below 20nm. In a recent work \cite{KXscaling}, a general analytic extension was provided to the SN barrier to include the effects produced by the emitter's curvature, and then used to extend the Fowler-Nordheim equation \cite{KXnonfn} and the GTF theory \cite{KXGTF} for emitters with radii down to 4-5nm. However, modern electron sources might be even sharper than this and a more general computational approach might be needed to calculate accurately the emission current and the Nottingham effect \cite{Nottingham,NottingCharb}.

For example, multi-physics simulations by means of combining Molecular Dynamics (MD) \cite{Parviainen2011, Eimre2015, Veske2016} or Kinetic Monte-Carlo (KMC) \cite{janssonKMC} with electrodynamics Finite Elements Method (FEM) or Finite Difference Methods (FDM) can be used to analyse the evolution of even atomically sharp metallic tips under strong electric fields. Since both MD and KMC approaches do not include the electron dynamics explicitly, they have to rely on the classical interpretation of emission currents given by either the standard FN or GTF equations to predict the emitted current,  in order to estimate the effect of the Joule heating of the tips. This approach overlooks two significant factors: 1) the contribution of the Nottingham effect to the heating or cooling of the emitters and 2) the fact that standard equations using the SN barrier might overestimate the current by several orders of magnitude when the emitter is of nanometric size. 

To tackle the mentioned problems, we have developed a general algorithm for the estimation of electron emission, which aims to calculate accurately  the emission currents and the Nottingham effect in addition to resistive heating for all three emission regimes (thermal, field and intermediate). Our method takes into account the possible high curvature of the emitters, since it does not assume a specific tunnelling barrier shape, which in general depends on the specific geometry of the emitter. The algorithm automatically distinguishes between different emission conditions and uses the appropriate analytical approximations if applicable, or numerically evaluates the JWKB formula if necessary. Thus it reduces the computational cost without losing the accuracy of the JWKB approach.

Presently we present the implementation of our method in the MD-FDM simulation \cite{Djurabekova2011,Parviainen2011} of the heat evolution of Cu nanotips. The usage of the algorithm makes it possible to take into account two new effects: 1) the Nottingham effect heat component on top of the resistive heating one and 2) the significant reduction of electron emission due to the high emitter curvature. The inclusion of these factors in the simulations leads to significantly different results on the thermal evolution of the sharp Cu nanotips.

The algorithm is realized in a Fortran 2003 computational tool, named "General Tool for Electron Emission Calculations - GETELEC". GETELEC aims to provide an accurate, computationally inexpensive and easy-to-use tool for the estimation of electron emission under various conditions and shapes of emitters. The tool is fully open-source and can be easily integrated in existing simulation software, but also be used separately for general purpose calculations. The source code is available in \citep{GETELEC} along with the corresponding documenta-
tion.

The paper is organized as following. In the method section we describe the basic ideas realized in the algorithm of the proposed tool and its implementation. First we propose a general model for the tunnelling barrier and then we explain the effect of the sharpness of field emitting tips. Consequently we give details on the electron emission calculation in the three different regimes, and finally we describe how our algorithm is utilized in the simulation of the thermal evolution of Cu nanotips. In the results section we show the results of electron emission calculations with GETELEC. First we present some general testing calculation results. Then we present the results of MD-FDM simulations for the thermal evolution of Cu nanotips and discuss their difference compared to the FN or GTF models. Finally we show that GETELEC can be successfully used to easily analyse experimental field emission measurements. 

\section{Method}

\subsection{A generalized shape of the tunnelling barrier}
\label{sec:model}

In order to calculate the electron emission current density from a metal, there are three input parameters that must be specified. The work function $\phi$, the local temperature $T$ and the shape of the electrostatic potential $V(x)$ in the vicinity of the emission point. Here $x$ stands for the distance from the emission point in the vacuum along the most probable path in the quantum mechanical sense \cite{Kapur}. Throughout this paper we will assume that the latter path coincides with the straight-line path perpendicular to the emitting surface, which is a good approximation when only the value of the total current is of interest and not the distribution of the electron beam \cite{KXBeamspot}.
 
It is the assumptions on the shape of $V(x)$ that limit the validity of the FN and GTF equations and in some cases, even their recent extensions of ref. \cite{KXnonfn, KXGTF}. In the standard theories based on the SN barrier, the electrostatic potential is approximated to be linear $V(x)=Fx$ where $F$ is the local field at the emitting point. The latter approximation is in principle valid only for planar surfaces, and practically valid for emitter radii of curvature $R$ grater than 15-20nm. In the recent extensions of \citep{KXnonfn,KXGTF}, a quadratic curvature-dependent term was added to the linear one, giving the form
\begin{equation}
    \label{eq:Vquad}
    V(x)=Fx-\frac{F}{R}x^2
\end{equation} 
which extends the validity of the approximation down to radii of 4-5nm. In general, $V(x)$ has an arbitrary shape that depends on the whole emitter geometry. Our algorithm gives the option to input this shape directly as a data array  $(x_i,V_i)$ in case it has already been calculated from the solution of the Laplace equation by other means (e.g. FEM or FDM). 

In case such a calculation is not available, $V(x)$ can be approximated by a general model with three adjustable parameters $(F,R,\gamma)$ which describe $V(x)$ as: 
\begin{equation}
    \label{eq:model}
    V(x) = F\frac{R(\gamma-1)x+x^2}{\gamma x+R(\gamma-1)}.
\end{equation}   
These parameters are:
\begin{equation}
    \label{eq:params}
    F = V'(0),\ R = \frac{-2F}{V''(0)},\ \gamma = \frac{F}{V'(x \rightarrow \infty)}.
\end{equation}
Note that the above relations are valid for eq. \eqref{eq:model}. The expression for $R$ is not obvious, but it was proved to be general in ref. \cite{KXnonfn}, and the quadratic approximation of eq. \eqref{eq:Vquad} follows from it.   

It was found after multiple tests that in most of the cases, the potential near sharp emitters of various geometries, calculated by various methods (FDM, FEM, point-matching), can be fitted accurately to the above model with appropriate selection of $(F,R,\gamma)$. We note that simpler potential models such as the linear SN barrier or the barrier with the quadratic extension term of eq. \eqref{eq:Vquad} are special cases of eq. \eqref{eq:model} for $\gamma=1$ and $\gamma=0$ respectively. Note that when $\gamma < 1$, the model of eq. \eqref{eq:model} has physical meaning only in the region close to $x=0$. For example, in the case of $\gamma = 0$, eq. \eqref{eq:model} collapses to \eqref{eq:Vquad} and the far field $V'(x \rightarrow \infty)$ goes to $-\infty$. However it remains a generally valid approximation for $x \ll R$, as proved in \cite{KXnonfn}.  
\begin{figure}[htbp]
    \centering
    \includegraphics[width=\linewidth]{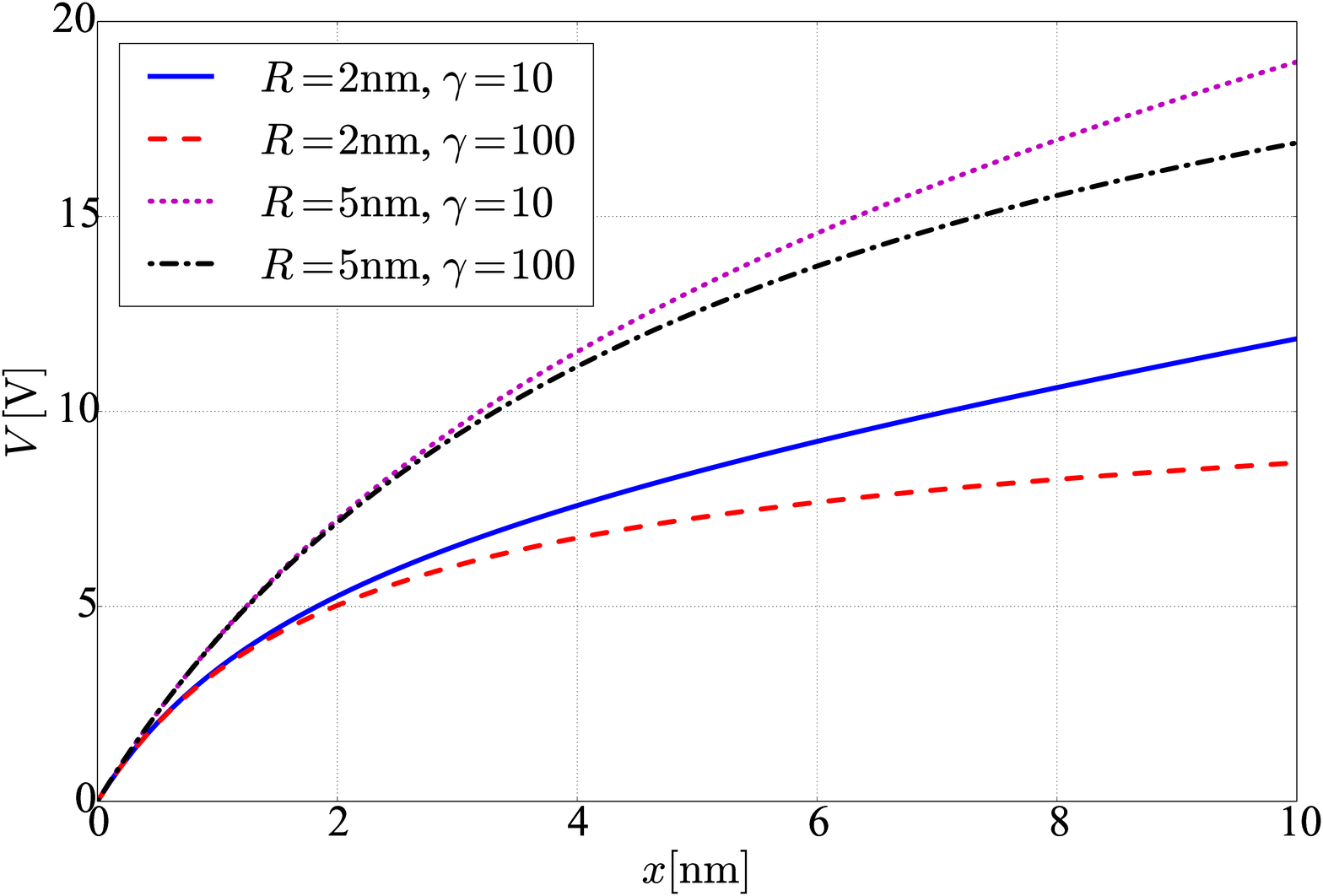}
    \caption{$V(x)$ according to eq. \eqref{eq:model} for various $(R,\gamma)$.}
    \label{fig:model}
\end{figure}

In figure \ref{fig:model} we show the behaviour of the potential model for different values of $(R,\gamma)$ (variation of $F$ does not present any interest as it is only a multiplicative scaling factor). We can see that the shape of $V(x)$ depends much more on $R$ than on $\gamma$, especially in the region close to the surface ($x=0$), which is usually of the highest interest. On the other hand, in the far region the potential has a linear form with a slope that depends only on $\gamma$.

In order to obtain the total barrier potential energy $U(x)$, we have to add the work function $\phi$ and the image interaction. For the latter we use the standard expression for the spherical image potential \cite{JensenImage}. We obtain
\begin{equation}
    U(x)=\phi-eV(x)-\frac{Q}{x(1+x/2R)}
    \label{eq:Uofx}
\end{equation}
where, $Q=e^2/16\pi\epsilon_0\approx0.36e$V nm is the standard image pre-factor and $e$ is the elementary charge. We have also taken the reference energy level to be the Fermi energy, i.e. $E_F \equiv 0$.

Finally, the shape of the electrostatic potential $V(x)$ can be introduced in the calculations by reading either an array $(x_i, V_i)$ or the three parameters $(F,R,\gamma)$ of the model \eqref{fig:model} . In the first case, there are three options for the evaluation of the smooth function $V(x)$ from the discrete points $(x_i, V_i)$: 1) spline interpolation, 2) fitting a polynomial of sufficient degree or 3) fitting the model \eqref{eq:model} to the data by using the Levenberg-Marquardt algorithm \cite{Levenberg,Marquardt}. Sometimes evaluation of a polynomial or eq. \eqref{fig:model} (options 2 and 3) is more efficient than spline interpolation (option 1). In such cases, our algorithm attempts fitting the input data first and if the attempt is unsatisfactory, returns automatically to the default option 1.

\subsection{"Sharp" and "Blunt" emitters}
\label{S-B barriers}

In the proposed algorithm, it is important to make a choice of the appropriate approximation suitable for a given case. The first decision that the algorithm makes, is how the tunnelling transmission coefficient $D$ will be calculated. The choice is made automatically based on the input data. It is important to define whether the tip is "sharp" (nanoscale) or "blunt" (macroscopic scale). If the tip is "sharp", in some cases even the approximations suggested in \cite{KXnonfn,KXGTF} may not be sufficient.

The transmission coefficient is given within the JWKB approximation by the Kemble formula \cite{Kemble}
\begin{equation}
    D(E)= \frac{1}{1+\exp\left(G(E)\right)} \textrm{,}
    \label{eq:Kemble}
\end{equation}
where $G(E)$ is the Gamow exponent. $G$ depends on the whole shape of the tunnelling barrier $U(x)$ and is obtained by the JWKB integral
\begin{equation}
    G(E)=g \int_{x_1}^{x_2}\!\sqrt{\left(U(x)-E\right)}\,dx \textrm{,}
    \label{eq:Gamow}
\end{equation}
where $E$ is the electron's energy, $x_1$ and $x_2$ are the turning points where $U(x)=E$ and $g=\sqrt{8m}/ \hbar \approx 10.246 (e\textrm{V})^{-1/2} (\textrm{nm})^{-1}$.
In the case of "blunt" emitters, $G$ will be calculated according to the analytic approximation of eq. (11a) of ref. \cite{KXGTF}. On the other hand, if the emitter is considered "sharp", eq. \eqref{eq:Gamow} will be integrated numerically.

\begin{figure}[htbp]
    \centering
    \includegraphics[width=\linewidth]{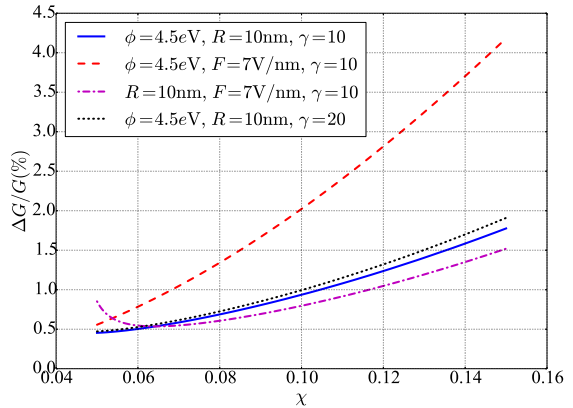}
    \caption{Relative error in the calculation of $G$ by the "blunt" approximation as a function of $\chi$. The four curves correspond to different combinations of parameters that produce the same $\chi$. Without loss of generality we assume $E=0$. }
    \label{fig:Gerror}
\end{figure}

The sharpness of the emitter is defined by the sharpness parameter $\chi \equiv (\phi-E)/eFR$. In ref. \cite{KXGTF} it was shown that the (absolute) error of the quadratic approximation for all calculated parameters goes to zero as $O(\chi^2)$ when $\chi \ll 1$. This is shown numerically in figure \ref{fig:Gerror} where we plot the relative error in the calculation of $G$ when it is obtained by the analytic approximation in comparison to its full numerical calculation as a function of $\chi$. We show the behaviour of the error for four different sets of parameters. We see that the error depends strongly on $\chi$ and it goes to zero when $\chi \rightarrow 0$ (except in the case of varying $\phi$, where $G$ goes to zero as well as $\Delta G$).  

If we require the relative error in the calculation of $G$ to be less than, say, $2\%$, we demand that $\chi<\chi_{max}=0.1$ for the approximate formulas to be used. $\chi_{max}$ can be further optimized to obtain a reasonable balance between high accuracy of calculations and computational efficiency.

After the "sharpness" of the barrier is defined and the method for obtaining $G$ is selected, the algorithm calculates four crucial quantities: 1) the maximum of the potential barrier $U_m=\max\{U(x)\}$, 2) $n_F = \left[-k_BTG'(0) \right]^{-1}$, 3) $n_T = \left[-k_BTG'(U_m) \right]^{-1}$ and 4) $G(0)$. Here $k_B$ is the Boltzmann constant and $T$ is the temperature. All those quantities are calculated numerically when $\chi>\chi_{max}$ and analytically according to the approximations developed in ref. \cite{KXGTF} when $\chi<\chi_{max}$.

\subsection{Field, Thermal and Intermediate regimes}
\label{sec:regimes}

The second decision that the algorithm makes is how to approach the integration over energies needed to calculate the current density \cite{MurphyG} 
\begin{equation}
    J= Z_S k_BT \int_{-\infty}^{\infty}~D(E)\log\left(1+\exp\left(-E/k_BT\right)\right)dE
    \label{eq:Jint}
\end{equation}
and the Nottingham heat deposited in the emitter per unit area \cite{Jensen2008}
\begin{equation}
    P_N = Z_S \int_{-\infty}^{\infty}~ \frac{E}{1+\exp (E/k_BT)} \int_{-\infty}^{E}~ D(\epsilon) d\epsilon dE .
    \label{eq:Pint}
\end{equation}
In the above equations, $Z_S \equiv em/2\pi^2\hbar^3 \approx 1.618 \times 10^{-4} A(eVnm)^{-2}$ is the Sommerfeld current constant \cite{Sommerfeld}.
This decision is made according to Jensen's theory \cite{Jensen2008} which gives the appropriate separation criteria among thermal, field and intermediate regimes. 

Jensen showed \cite{Jensen2006} that the two parameters $n_F$ and $n_T$, defined in section \ref{S-B barriers}, can be used to define whether the electron current originates from field or thermionic emission. When $n_F \gg 1$ the thermionic component of the emission current is negligible and when $n_T \ll 1$ the emission is purely thermionic.   In those cases, equations (1) and (19)-(20) of ref. \cite{Jensen2008} give good approximations for $J$ and $P_N$. We note here a typographical mistake in eq. (20) of ref. \cite{Jensen2008}: the exponent of the first term should be "s" and not "n". 

We employ this approach to determine in which regime the electrons are emitted from the tip. We introduce two numbers $n_{high}$ and $n_{low}$ to separate two regimes: field emission is considered if $n_F > n_{high}$ and emission is treated as thermionic if $n_T < n_{low}$. In both cases, the approximate formulas are used. Otherwise, electron emission must be considered in the intermediate regime, $G(E)$ has to be calculated for the whole energy spectrum and the expressions \eqref{eq:Jint} and \eqref{eq:Pint} are integrated numerically. Note that $G(E)$ might be calculated by different methods at different energies since $\chi$ might cross $\chi_{max}$ as $E$ increases.

The limits $(n_{high},n_{low})$ depend on the required accuracy and can be adjusted prior to calculations. For example, $n_{low}=0.6$ and $n_{high}=2.3$ result in satisfactory accuracy with an error (comparing with full numerical calculation) of less than 2.5\% in the value of $J$. 

However, the algorithm gives the option to forcefully use the GTF approximations of ref. \cite{Jensen2008} in the intermediate regime if only a rough estimation electron emission is required. In that case $(n_{high},n_{low})$ are set to 1.

\subsection{Simulation of the thermal evolution of Cu nanotips}

GETELEC was implemented and tested with the existing MD-FDM code HELMOD. HELMOD \cite{Djurabekova2011} concurrently couples the MD code PARCAS \cite{GhalyParcas,NordlundParcas} with the solution of the Laplace equation by using the FDM. This code was previously used to study the thermal evolution \cite{Parviainen2011,Eimre2015} and stability \cite{Veske2016} of Cu nanotips under strong electric fields.

Up to now, HELMOD was relying on the classical FN \cite{Parviainen2011} or GTF \cite{Eimre2015,Veske2016} equations for the calculation of electron emission currents and the Joule heating effect was considered to be the main heat source. Here we shall use GETELEC to calculate the electron emission heating effects, including the Nottingham heating component and taking into account the very high curvature of the nanotips. 

At each MD time step $\Delta t = 4.05\textrm{fs}$, HELMOD solves the Laplace equation and obtains the 3D electrostatic potential $V(\vec{r})$ by dividing the whole space into the FDM grid $(\delta x, \delta y, \delta z)$. Then $V(\vec{r})$ along with the previous step temperature distribution $T_{old}$ are passed to GETELEC which calculates the current density $J$ and the Nottingham heat deposited per unit area $P_N$ at each grid point that belongs to the metal-vacuum surface.

In line with ref. \cite{Parviainen2011}, we will consider the thermal distribution along the emitter as one-dimensional, because the height of the nanotip is much greater than its other dimensions. The tip is thus considered as a stack of "slices" $\delta z$. Each slice is crossed by a total current $I(z_i)$, has a number $N_i$ of surface points and a cross section $A_i$. Each $j$-th surface point of the $i$-th slice emits a current density $J_j(z_i)$ and produces a Nottingham heat $P_{Nj}(z_i)$, that are obtained by GETELEC. The slices have the width of a crystal monolayer.  The total current is then given by:
\begin{equation}
    \label{eq:Ii}
    I(z_i) = I(z_{i-1}) + \sum_{j=1}^{N_i}J_j(z_i)\delta A
\end{equation}
where $\delta A$ is the elementary area of each surface grid point and the $i$-counting goes from the top to the bottom of the tip. The total heat deposited on each slice is then found by
\begin{eqnarray}
    \label{eq:Pi}
    P(z_i) = P_J(z_i) + P_N(z_i) = \nonumber \\
     = I^2(z_i) \frac{\rho \delta z}{A_i}  + \sum_{i=1}^{N_i}P_{Nj}(z_i)\delta A
\end{eqnarray}
where $\rho$ is the resistivity of Cu (multiplied by the finite size correction factor).

Then we insert $P(z_i)$ along with the previous temperature distribution $T_{old}(z_i)$ in the heat equation introduced in ref. \cite{Parviainen2011} and by solving it for the time interval of one MD time step, we obtain the next temperature distribution $T_{new}(z_i)$. The latter is inserted to the MD lattice by using the Berendsen \citep{Berendsen} temperature control scheme with $\tau=0.5ps$ for each slice. The heating process is very slow and it is a good approximation to assume that the electrons are in thermal equilibrium with the lattice. The described procedure is continued repeatedly for many time steps.

\section{Results}

\subsection{$J$ and $P_N$ in different regimes}

\begin{figure}[htbp]--
    \centering
    \subfloat{\includegraphics[width=\linewidth]{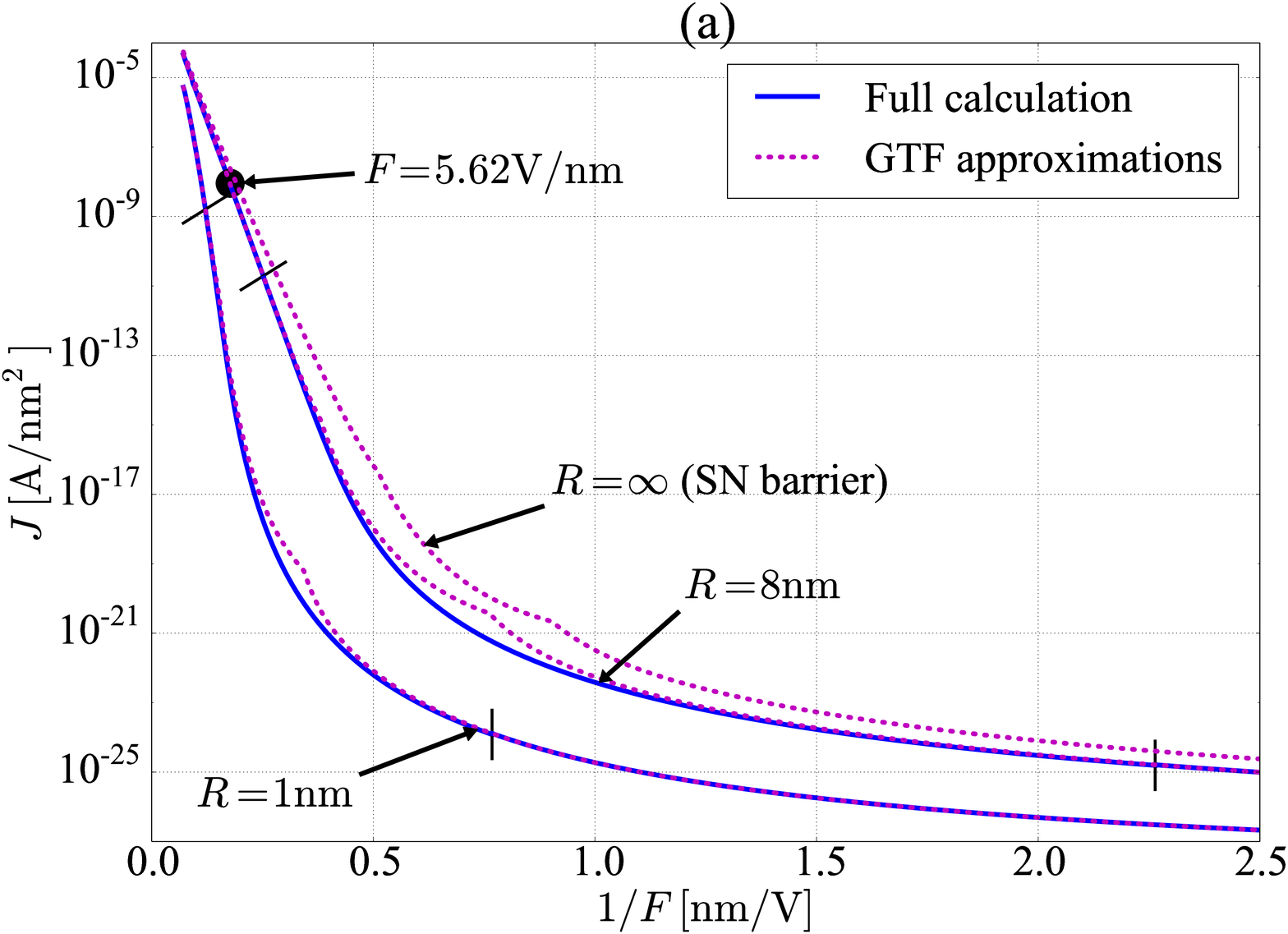}}

    \subfloat{\includegraphics[width=\linewidth]{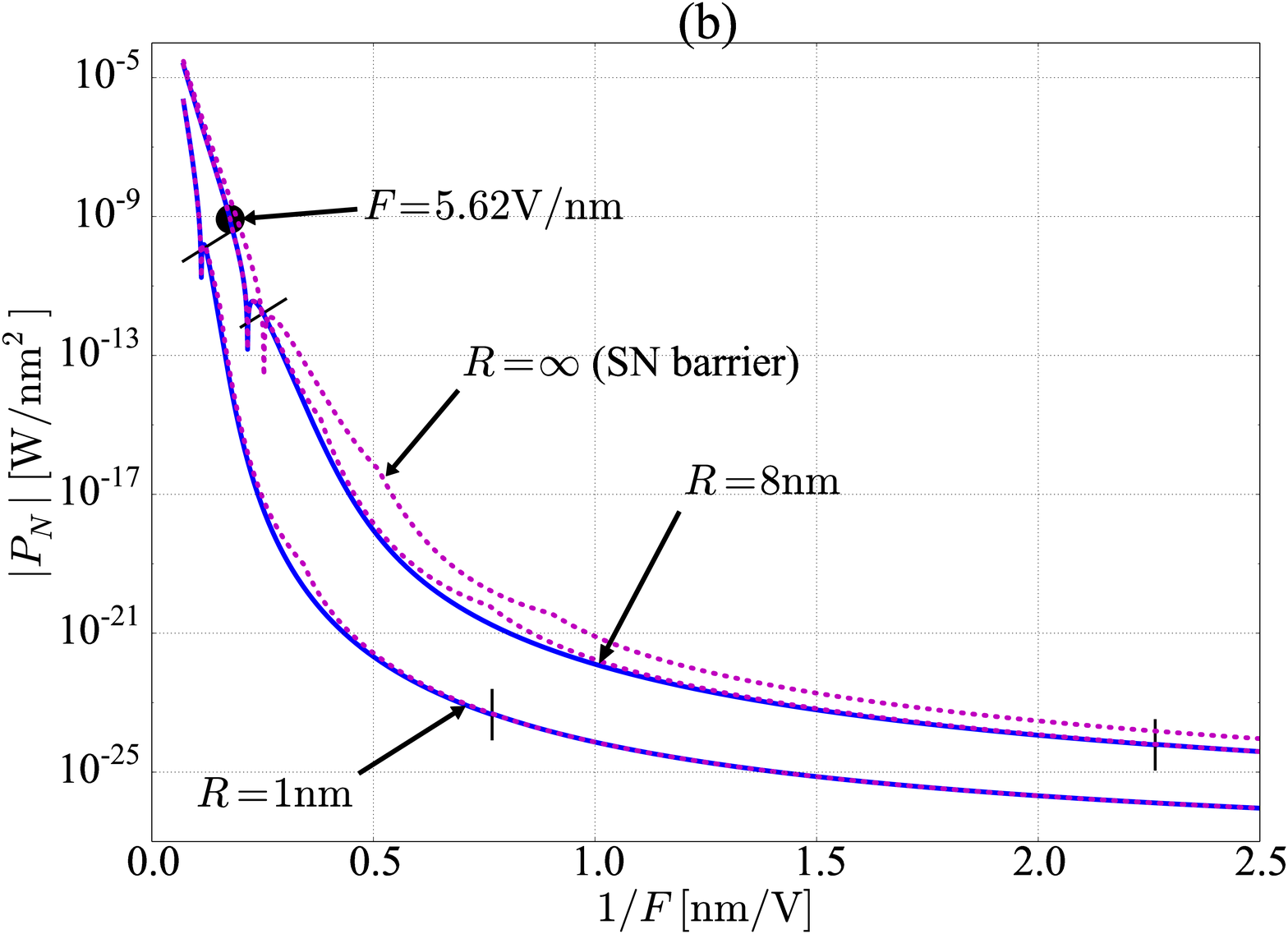}}
    
    \caption{$J$ and $|P_N|$ as a function of $F$ for various $R$ as calculated by our algorithm. Solid blue lines correspond to full calculation with numerical integration of equations \eqref{eq:Jint}, \eqref{eq:Pint} in the intermediate regime while dotted magenta ones correspond to usage of the GTF approximations. The separation bars indicate the turning points between the three emission regimes, and the circle marker in the top-left side the turning point of $F=5.62\textrm{V/nm}$ between "blunt" and "sharp" approximations for $R=8\textrm{nm}$. The downwards spikes in figure (b) correspond to 0 crossings, where Nottingham heating in the high-field region turns to cooling as the field decreases.}
    
    \label{fig:JandP}
\end{figure}

In order to validate the electron emission calculation methods described in sections \ref{sec:model}-\ref{sec:regimes}, we plot $J$ and $P_N$ versus the local electric field $F$ for various radii $R$, as obtained by using the proposed algorithm with $T=1000K$, $\phi=4.5eV$ and $\gamma=15$ (see figure \ref{fig:JandP}). The temperature rise was not considered in this case. Solid blue lines correspond to full calculation with numerical integration of eq. \eqref{eq:Jint}, \eqref{eq:Pint} in the intermediate regime, while dotted magenta lines correspond to implementation of the approximative GTF formulas. The bar lines indicate the turning points where the emission regime changes from field (left - high field region) to intermediate (middle field region) and thermal (right - low field region). The marker in the top-left side where $F=5.62\textrm{V/nm}$ indicates the point where the "blunt" approximations start being used as the field increases in the curve for $R=8nm$.

We see that when our algorithm is fully used (solid blue lines), the curves are almost perfectly continuous and matched. This ensures that different approximations agree with each other at their turning points and are asymptotically valid as theoretically expected. 

Moreover, the GTF approximations for the energy integrals (dotted magenta lines) produce some errors of a factor of 8.5 at maximum. However, not taking into account the curvature of the barrier (assuming the SN barrier - $R=\infty$), may produce extreme errors in $J$ and $P_N$ of up to 8 orders of magnitude for nanoemitters with $R=1nm$ (compare the solid line with the dotted corresponding to $R=\infty$). Also the positions of the turning points between different regimes differ significantly for various $R$. This is indicative of the importance of the emitter curvature on the shape of the barrier and all the emission characteristics.

Finally, we note that for $R=8\textrm{nm}$, the "blunt" approximation is valid for $F>5.62\textrm{V/nm}$ (upper-left side of the circle marker in figure \ref{fig:JandP}). This shows that in certain cases the "blunt" approximation can be valuable since it significantly reduces the computational cost. Here we have used the criterion of validity described in section \ref{S-B barriers} $((\phi-E)/eFR)<0.1$ which is relatively strict.   

In view of the above, it becomes clear that usage of GETELEC becomes necessary when emitters are highly curved and even the approximations for curved emitters of ref. \cite{KXGTF} fail.

\subsection{Thermal evolution of Cu nanotips}

In order to show the impact of the usage of GETELEC in MD-FDM dynamic simulations, we simulate the evolution of a cylindrical nanotip with hemispherical top of radius $R=1.5\textrm{nm}$ and height $h=20.3\textrm{nm}$, in a rectangular simulation cell of $28.9\textrm{nm} \times 28.9\textrm{nm} \times 28.9\textrm{nm}$. The initial temperature of the nanotip, which is also equal to the boundary condition of the bulk temperature was set to $T_0=300\textrm{K}$. We use the standard work function for Cu [100] $\phi=4.7e\textrm{V}$ and an applied macroscopic field of $F_{mac} = 1.25\textrm{V}/ \textrm{nm}$. Note that as in ref. \cite{Parviainen2011}, we have taken into account the finite size effect for the nanotips, and we have multiplied the electric and thermal resistivity of Cu by a correction factor, which in the case of $R=1.5\textrm{nm}$ is 20.74 \cite{FSE2016}. 

We run simulations in two different modes. In mode (A) we calculate the heating by taking into account both Joule and Nottingham effects as calculated by GETELEC. In mode (B) we include only the Joule heat produced by the current which is calculated by the GTF equation, as it was done in the previous publications. 

\begin{figure}[htbp]
    \centering
    \includegraphics[width=\linewidth]{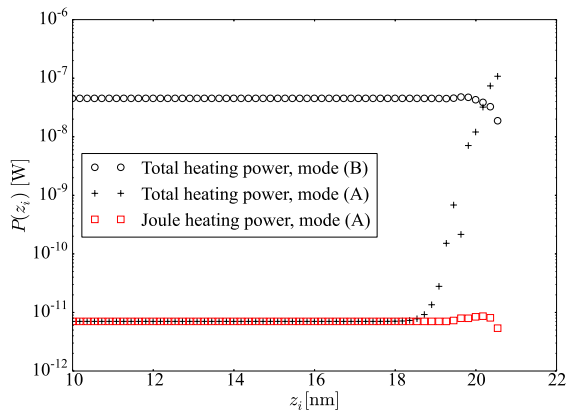}
    \caption{Total deposited heating power per "slice" $\delta z$ along the tip as calculated in mode (A) (crosses) and in mode (B) (circles). The Joule component of mode (A) calculation is also shown (red squares).}
    \label{fig:heats}
\end{figure}

In figure \ref{fig:heats} we compare the distributions of the initial (first time step after relaxation) deposited heat as calculated by mode (A) (crosses) and by mode (B) (circles). We also plot the Joule heat component of the mode (A) calculation (red squares). We see that the total current and hence the Joule heat are overestimated by several orders of magnitude from the GTF equation. Note that in the main body of the emitter, the Joule heat dominates because the emission and the Nottingham effect on the surface are practically zero, while the current flows through the bulk all along the emitter to supply the top. 

However, at the top of the emitter where most of the emission takes place, the Nottingham heat is about 5 orders of magnitude higher than the one produced by the resistive heating effect, thus playing more significant role in the heating process. This makes the two heat distributions be also qualitatively different. In mode (A), the tip is practically heated only at its top, while in the mode (B) it is heated almost uniformly. 

Since the radius of curvature of the nanotip is rather small, the GTF equation is expected to overestimate the current density. Hence we anticipate that in mode (A) the temperature distribution after some time evolution will give much higher temperature than in mode (B). Indeed, after a full evolution of the system for $40\textrm{ps}$, the emitter had already reached the melting temperature of $1358\textrm{K}$ at the top of the tip for the mode (B), while for (A) it had barely heated $37\textrm{K}$ above $T_0=300\textrm{K}$. It is evident that a significantly higher critical applied field is required to heat a nanotip than the one predicted by using the GTF equation.

\begin{figure}[htbp]
    \centering
    \includegraphics[width=\linewidth]{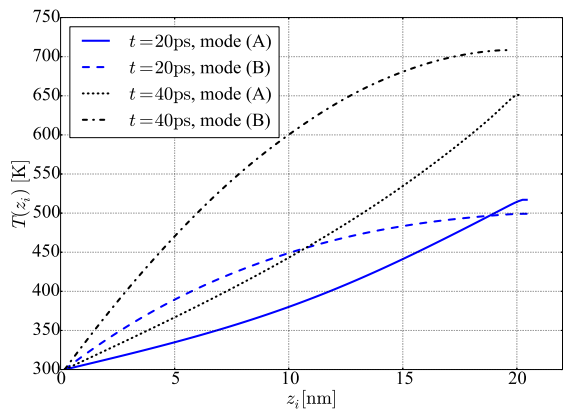}
    \caption{Resulting temperature distribution along the tip after 20ps (blue lines) and 40ps (black lines) of heating, as calculated by GETELEC including Nottingham effect  with $F_{mac}=1.5\textrm{V/nm}$ (solid and dotted lines) and by the GTF with only Joule heating for $F_{mac}=1.2\textrm{V/nm}$ (dashed and dashed-dotted lines).}
    \label{fig:temps}
\end{figure}
 
In figure \ref{fig:temps} we show the temperature distribution on the nanotip after a time interval of $20\textrm{ps}$ and $40\textrm{ps}$, again as calculated by mode (A) for an applied field of $F_{mac}=1.5\textrm{V/nm}$ and by mode (B) for $F_{mac} = 1.2\textrm{V/nm}$. We see that although the applied field in (A) is significantly higher, the maximum temperature at the top of the tip is lower than in (B). Note also, that the total current in the two cases is much different: $I \approx 4.6 \mu \textrm{A}$ for (A) and $I \approx 50\mu \textrm{A}$ for (B). This shows that the Nottingham effect is very significant in the heating process, since its inclusion makes the tip heat up to about the same temperature at the top, with 10 times lower current.

Furthermore, the shapes of the two distributions are different. The Nottingham effect is a surface phenomenon which dominates only at the top of the tip, thus producing a ramp form for the temperature distribution (practically the impulse response of the heat equation). On the other hand, the Joule effect is a bulk phenomenon depositing heat all along the tip. Therefore it produces a temperature distribution very close to the secant function predicted in ref. \cite{Parviainen2011} for mode (B). This might lead to melting and deformation only on the top of the tip for (A). This difference needs further investigation, as it might play a role in the possible initiation of a vacuum breakdown process.

We note finally that the usage of GETELEC instead of the GTF equation does not increase significantly the computational time required for carrying out the simulations. 

\subsection{Analysis of experimental I-V data}

Before concluding, we shall demonstrate how GETELEC can be used in general field emission calculations. One of the most frequent utilizations of field emission theory, is the analysis of experimental current-voltage $(I-V)$ measurements from field emitters. The usual procedure is that the data are plotted in the F-N plot form ($1/V - \log(I/V^2)$) and they are fitted to a straight line to reproduce the F-N equation and extract the field conversion factor $\beta$ (known also as "enhancement factor"). However, when the emitters are of nanometric size, or when the temperatures are high, the FN equation does not accurately describe the emission characteristics, and analysing experimental measurements becomes rather complicated.
  
GETELEC can tackle this problem since its results can be fitted to experimental $I-V$ data. We implemented the standard non-linear optimization algorithm "Trust Region Reflective" \cite{BranchTRF} in order to find the set of parameters $(F,\phi,R,T,\gamma)$ that best reproduces those data. We include the computational tool for this procedure in the GETELEC package (see the supplementary material) with the purpose of making such an analysis simple.

\begin{table*}[htbp]
    \centering
    \caption{Optimal parameters calculated by the fitting algorithm for 4 different data sets. $\phi$ was given fixed values. $\beta_{exp}$ stands for the $\beta$ obtained by other experimental methods. For data sets (1) and (3), $\sigma A_{eff}$ has been divided by the number of emitters in the corresponding array.}
    \label{tab:tab1}    
    \begin{tabular*}{\textwidth}[t]{l @{\extracolsep{\fill}} c @{\extracolsep{\fill}} c @{\extracolsep{\fill}} c @{\extracolsep{\fill}} c @{\extracolsep{\fill}} r}
    \hline \hline
    Data set & $\phi$               & $\beta$                       & $\sigma A_{eff}$                  & $R$                   & $\gamma$  \\
    \hline
    $1$         & $4.05e\textrm{V}$     & $0.017 \textrm{nm}^{-1}$      & $1.16\times 10^{-6}\textrm{nm}^2$ & $10.08\textrm{nm}$        & $9$       \\
    $2$         & $4.5e\textrm{V}$      & $1.007\beta_{exp}$            & $8.2\textrm{nm}^2$                    & $6.87\textrm{nm}$     & $100$     \\
    $3$         & $4.35e\textrm{V}$         & $0.065 \textrm{nm}^{-1}$      & $1.788\textrm{nm}^2$              & $16.22\textrm{nm}$        & $8.5$     \\
    $4$         & $4.5e\textrm{V}$      & $68.6$                            & $6.07\textrm{nm}^2$               & $2.96\textrm{nm}$     & $100$     \\
    \hline
    \end{tabular*}
\end{table*} 

It is standard in field emission theory to associate the local field $F$ with a macroscopic quantity $X$, which is either the macroscopic field $F_{mac}$ or the applied voltage $V_{appl}$, via $\beta \equiv F / X$. Thus the parameter that we really fit is $\beta$ and not $F$, since the latter is varied. Furthermore, an equivalent factor is needed to convert the microscopic current density $J$ into the macroscopic current $I$. The conversion factor is the "notional area" $\sigma A_{eff}$, where the effective area $A_{eff}$ accounts for the integration over the emitting surface and $\sigma$ is a correction pre-factor \cite{Modinos2001}.

Therefore, we define the free fitting parameters as $(\sigma A_{eff}, \beta, \phi, R, T, \gamma)$. Our fitting algorithm gives the option to set limits for any of those parameters. In cases where some of these parameters are known, it makes it easier and more reliable to fix them. For example, when the measurements are done in the cold field emission regime from an emitter with known work function, one can fix $\phi$ and $T$.
 
For testing and benchmarking purposes, we used GETELEC to fit the same experimental data used similarly in the previous publication \cite{KXnonfn}: data set (1) is from ref. \cite{Guerrera}, data set (2) is from ref. \cite{CabreraPRB} and (3) is from ref. \cite{Spindt2010}. In data set (4), we also include another set of measurements (measurement "S22" from ref. \cite{CERN2004}) to show that our algorithm extends to lower radii than the previous analytical approximations of ref. \cite{KXnonfn}. 

\begin{figure}[htbp]
    \centering
    \includegraphics[width=\linewidth]{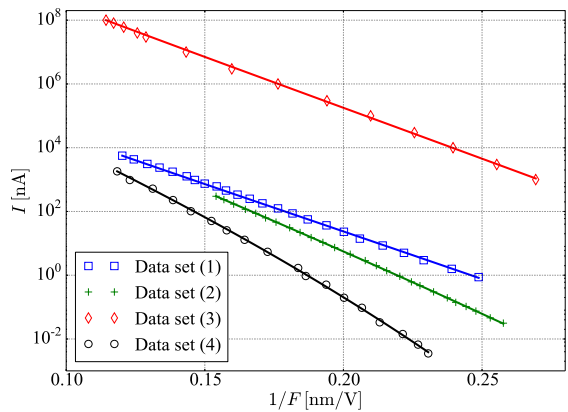}
    \caption{Experimental $I - F$ data (markers) from four different experimental groups. The solid lines are the corresponding theoretical fittings by GETELEC. Note that the local field $F$ in the horizontal axis is found by multiplying the experimental quantity $X$ by the extracted conversion factor $\beta$.}
    \label{fig:fittings}
\end{figure}

In figure \ref{fig:fittings} we show the $I-F$ plot of the experimental measurements (markers) along with the corresponding GETELEC calculations with the optimal parameters (solid lines). The extracted optimal parameters are given in table \ref{tab:tab1}. For the fitting we fixed $\phi$ to the value given at each corresponding experimental paper, and the temperature to $300K$. $R$ was limited within $1\textrm{nm}$ and $50\textrm{nm}$, $\gamma$ between $1$ and $100$ and $\beta$ was bounded so that $1\textrm{V/nm}<F<14\textrm{V/nm}$. $\sigma A_{eff}$ was completely free. We perform the calculations at room temperature, since all the field emission measurements are performed in the "stable operation" regime of the emitters, where the current densities are too low to produce any significant heating. 

Figure \ref{fig:fittings} shows that the model used in the proposed algorithm describes all the sets of experimental measurements with good accuracy. It also gives the ability to analyse  very easily $I-V$ data and extract desired parameters - especially $\beta$ and $R$ that might be of great interest from an experimental point of view. The optimal parameters we obtained in table \ref{tab:tab1} are very close to the previous ones of ref. \cite{KXnonfn}. However, there are some differences that can be attributed to the more accurate calculation realized in GETELEC, and to the different fitting technique. We also point out that the data set 4 has an $R$ which goes much lower than the limit of 5nm for the analytical approximations of ref. \cite{KXnonfn}.

It is worth noting that the the "enhancement factor" $\gamma$ used in the model (eq.(2)) is not related directly to the value $\beta$ commonly used to estimate the local enhancement of the external macroscopic electric field. The extracted $\gamma$ gives information on slight changes on the shape of the barrier, and not the actual conversion from macroscopic field to the local one. That information can be reliably taken only from $\beta$. 

\section{\label{sec:conc}Conclusions}
In conclusion, we have developed an algorithm and computational tool for electron emission calculations, valid for nanometric size emitters. Our algorithm covers all the regimes of the general thermal-field emission and uses either analytical equations when applicable, or numerical integration of the JWKB approximation to calculate the current density and the Nottingham heat. The latter can be integrated in the heat diffusion equation to analyse the heat distribution due to its combination with resistive heating. 

We have implemented our algorithm in MD simulations of the thermal evolution of Cu nanotips, and we have found that taking into account the curvature of the emitters and the Nottingham effect produces significantly different results in comparison with using the standard Fowler-Nordheim or General Thermal-Field equations for calculating electron emission. The suggested algorithm can also be used to analyse experimental current-voltage data and extract the enhancement factor and the radius of curvature of an emitter.

\section*{Acknowledgements}

The authors acknowledge gratefully the financial support by the Academy of Finland (Grant No. 1269696). A. Kyritsakis would like to thank Dr. Stefan Parviainen for his valuable help in the integration of  GETELEC with HELMOD.

\section*{References}

\bibliography{bibliography}

\end{document}